\documentclass[journal]{IEEEtran}
\usepackage[left=0.67in,right=0.67in,top=0.66in,bottom=0.67in]{geometry}
\usepackage[T1]{fontenc}
\usepackage[latin9]{inputenc}
\usepackage{amsthm}
\usepackage{amsmath} 
\usepackage{graphicx} 
\usepackage{color} 
\usepackage{cite}
\usepackage{algpseudocode}
\usepackage{algorithm}
\usepackage{amssymb}
\usepackage{subfigure}
\usepackage{esint}
\usepackage{algpseudocode}
\usepackage{epstopdf}
\usepackage{multirow}
\usepackage{makecell}
\usepackage{float}
\usepackage{lipsum}
\usepackage{balance}

\newtheorem{definition}{Definition}
\newtheorem{remark}{Remark}

\newtheorem{example}{Example}

\theoremstyle{plain}
\theoremstyle{plain}
\newtheorem{theorem}{Theorem}

\newcommand{\comment}[1]{}

\IEEEoverridecommandlockouts

\allowdisplaybreaks

\begin{document}

\title{Managing Device Lifecycle: Reconfigurable Constrained Codes for M/T/Q/P-LC Flash Memories}

\author{
   \IEEEauthorblockN{Ahmed Hareedy, \IEEEmembership{Member, IEEE}, Beyza Dabak, and Robert Calderbank, \IEEEmembership{Fellow, IEEE}} \vspace{-1.0em}
   
   \thanks{A. Hareedy, B. Dabak, and R. Calderbank are with the Department of Electrical and Computer Engineering, Duke University, Durham, NC 27708 USA (e-mail: ahmed.hareedy@duke.edu; beyza.dabak@duke.edu; robert.calderbank@duke.edu). This research was supported by NSF under grant CCF 1717602.}
}
\maketitle

\begin{abstract}
Flash memory devices are winning the competition for storage density against magnetic recording devices. This outcome results from advances in physics that allow storage of more than one bit per cell, coupled with advances in signal processing that reduce the effect of physical instabilities. Constrained codes are used in storage to avoid problematic patterns, and thus prevent errors from happening. Recently, we introduced binary symmetric lexicographically-ordered constrained codes (LOCO codes) for data storage and data transmission. LOCO codes are capacity-achieving, simple, and can be easily reconfigured. This paper introduces simple constrained codes that support non-binary physical substrates; multi, triple, quad, and the currently-in-development penta-level cell (M/T/Q/P-LC) Flash memories. The new codes can be easily modified if problematic patterns change with time. These codes are designed to mitigate inter-cell interference, which is a critical source of error in Flash devices. The occurrence of errors is a consequence of parasitic capacitances in and across floating gate transistors, resulting in charge propagation from cells being programmed to the highest charge level to neighboring cells being programmed to lower levels. This asymmetric nature of error-prone patterns distinguishes Flash memories. The new codes are called $q$-ary asymmetric LOCO codes (QA-LOCO codes), and the construction subsumes codes previously designed for single-level cell (SLC) Flash devices (A-LOCO codes). QA-LOCO codes work for a Flash device with any number, $q$, of levels per cell. For $q \geq 4$, we show that QA-LOCO codes can achieve rates greater than $0.95 \log_2 q$ information bits per coded symbol. The complexity of encoding and decoding is modest, and reconfiguring a code is as easy as reprogramming an adder. Capacity-achieving rates, affordable encoding-decoding complexity, and ease of reconfigurability support the growing development of M/T/Q/P-LC Flash memory devices, as well as lifecycle management as the characteristics of these devices change with time, which significantly increases their lifetime.
\end{abstract}

\section{Introduction}\label{sec_intro}

Data storage densities are increasing rapidly as modern applications, e.g., internet of things (IoT) applications, access, process, and store more and more data. In 2015, the storage density of Flash memory devices surpassed that of magnetic recording (MR) devices. This milestone resulted from multiple advances in physics, architecture, and signal processing. The major advance in Flash physics was enabling more than two storage levels, i.e., more than two charge levels, per cell, and thus allowing the storage of more than one bit per cell. The major advance in Flash architecture was devising the three-dimensional vertical NAND Flash structure.

The data storage industry achieves high reliability by combining constrained codes, designed to avoid problematic patterns, with error-correcting codes (ECCs), designed to correct the errors that remain. Run-length-limited (RLL) codes are a class of constrained codes introduced in 1970 \cite{tang_bahl}, that were first used to improve the storage density of early MR devices employing peak detection \cite{siegel_mr, immink_surv}. Modern storage devices employ sequence estimation rather than peak detection, but constrained codes are still used to improve performance \cite{immink_surv, siegel_const}. RLL codes also find application in optical recording \cite{immink_opt}.  When first introduced in \cite{tang_bahl}, lexicographic indexing was used to encode and decode RLL codes, but this was replaced by methods based on finite-state machines (FSMs) in later work \cite{ach_fsm}. RLL codes are associated with transition-based signaling.

In level-based signaling, each symbol (or bit) is associated with a distinct level for storage or transmission. For example, in the binary case, a $0$ is represented by $A_0$ and a $1$ is represented by $A_1$, where $A_0 < A_1$, in what is called bipolar non-return-to-zero (NRZ) signaling. A binary symmetric $\mathcal{S}_x$-constrained code is a code that forbids the patterns in the set $\mathcal{S}_x \triangleq \{010, \allowbreak 101, 0\bold{1}^20, 1\bold{0}^21, \dots, 0\bold{1}^x0, 1\bold{0}^x1\}$ from appearing in any codeword, where the notation $\bold{y}^r$ refers to a sequence of $r$ consecutive $y$'s. A binary asymmetric $\mathcal{A}_x$-constrained code is a code that forbids the patterns in the set $\mathcal{A}_x \triangleq \{101, \allowbreak 1\bold{0}^21, \dots, 1\bold{0}^x1\}$ from appearing in any codeword. Both $\mathcal{S}_x$-constrained codes and $\mathcal{A}_x$-constrained codes are associated with level-based signaling, which is natural for Flash.

In Flash devices, inter-cell interference (ICI) is one of the main sources of errors. Parasitic capacitances in and across floating gate transistors result in charge propagation from cells being programmed to the highest charge level to neighboring cells being programmed to lower levels.\footnote{Asymmetric errors resulting from charge leakage and other problems in Flash devices are handled by error-correction techniques \cite{ahh_jsac, ahh_nboo}.} Thus, unintentional increases in charge values occur, resulting in errors during reading. The authors of \cite{qin_flash} and \cite{kayser_flash} introduced constrained codes to prevent the level pattern $(q-1)0(q-1)$ from being written in a Flash device with $q \geq  2$ levels per cell.\footnote{Note that charge levels directly translate to threshold voltage levels. For simplicity, levels are defined by their indices $\{0, 1, \dots, q-1\}$.} Via extensive experiments, the authors of \cite{veeresh_mlc} demonstrated that for multi-level cell (MLC) Flash devices ($4$ levels per cell), the set of level patterns to be forbidden (contribute the most to ICI) should be $\{303, 313, 323\}$. This set was recently generalized in \cite{chee_qlc} to $\{(q-1)0(q-1), (q-1)1(q-1), \dots, \allowbreak (q-1)(q-2)(q-1)\}$ for a Flash device with $q$ levels per cell.

In previous work \cite{ahh_loco}, we introduced capacity-achieving $\mathcal{S}_x$-constrained codes, named lexicographically-ordered $\mathcal{S}_x$-constrained codes (LOCO codes), that make significant MR density gains possible. LOCO codes are simple, and they can be easily reconfigured to support additional constraints. The $\mathcal{A}_x$-constraint forbids ICI-causing patterns in single-level cell (SLC) Flash devices ($2$ levels per cell). The advantage of designing codes for asymmetric errors, rather than symmetric errors, is that it becomes possible to achieve notably higher rates. In \cite{ahh_aloco}, we designed capacity-achieving $\mathcal{A}_x$-constrained codes, named asymmetric LOCO codes (A-LOCO codes), that offer a better rate-complexity trade-off than previous codes, and that can be easily reconfigured. We anticipate using a combination of machine learning and analysis of errors collected before the ECC decoder to identify new patterns that need to be forbidden as the device ages. We see (A-)LOCO codes as a method of extending device lifetime.

In this paper, we generalize our asymmetric constrained codes in \cite{ahh_aloco} to Flash devices with any number, $q$, of levels per cell. In particular, we introduce fixed-length $q$-ary asymmetric LOCO codes (QA-LOCO codes) for all Flash devices. QA-LOCO codes are capacity-achieving, and we devise the encoding-decoding rule for them to offer simplicity. While available literature only focuses on the effect of ICI on adjacent cells, we handle more general constraints for higher reliability in this work. QA-LOCO codes are also reconfigurable because of their encoding-decoding rule. We show that QA-LOCO codes can achieve significant lifetime gains for the Flash device with rates greater than $0.95 \log_2 q$ information bits per coded symbol, $q \geq 4$, at affordable complexities. Furthermore, we discuss ideas to reduce latency. We suggest that QA-LOCO codes can significantly improve the performance (increase the lifetime) of multi ($q=4$) and triple ($q=8$)-level cell Flash memories, and can remarkably accelerate the development of quad ($q=16$) and penta ($q=32$)-level cell Flash memories, which are the next generation.

The rest of the paper is organized as follows. In Section~\ref{sec_card}, we define QA-LOCO codes and introduce their cardinality. In Section~\ref{sec_rule}, we derive the QA-LOCO encoding-decoding rule. In Section~\ref{sec_rate}, we discuss rates and make comparisons. In Section~\ref{sec_algr}, we present the encoding and decoding algorithms and discuss reconfigurability. Section~\ref{sec_conc} concludes the paper

\section{Definition and Cardinality}\label{sec_card}

Denote a Galois field (GF) of size $q$ by GF($q$). Let $\alpha$ be a primitive element of GF($q$).\footnote{Our analysis works for any GF size $q$. However, we focus more on $q = 2^v$, $v \geq 1$, because of the nature of Flash devices. We write one symbol per cell.} Consequently,
\begin{equation}
\textup{GF}(q) \triangleq \{0, 1, \alpha, \alpha^2, \dots, \alpha^{q-2}\}. \nonumber
\end{equation}
We define $\delta$ as an element in GF($q$)$\setminus\{\alpha^{q-2}\}$ and also $\boldsymbol{\delta}^r \triangleq \delta_{r-1} \delta_{r-2} \dots \delta_0$ as a sequence in \big [GF($q$)$\setminus\{\alpha^{q-2}\}$\big ]$^r$. We now formally define QA-LOCO codes, which are $\mathcal{Q}^q_x$-constrained:

\begin{definition}\label{def_qaloco}
A QA-LOCO code $\mathcal{QC}^q_{m,x}$ with $q \geq 2$, $m \geq 1$, and $x \geq 1$ is defined by the following properties:
\begin{enumerate}
\item Each codeword $\bold{c}$ in $\mathcal{QC}^q_{m,x}$ has its symbols in GF($q$) and is of length $m$ symbols.
\item Codewords in $\mathcal{QC}^q_{m,x}$ are ordered lexicographically.
\item Each codeword $\bold{c}$ in $\mathcal{QC}^q_{m,x}$ does not contain any of the patterns in the set $\mathcal{Q}^q_x$, where:
\begin{equation}\label{eqn_setq}
\hspace{-1.14em}\mathcal{Q}^q_x \triangleq \{\alpha^{q-2}\delta\alpha^{q-2}, \alpha^{q-2}\boldsymbol{\delta}^2\alpha^{q-2}, \dots, \alpha^{q-2}\boldsymbol{\delta}^x\alpha^{q-2}\}.
\end{equation}
\item The code $\mathcal{QC}^q_{m,x}$ contains all codewords satisfying the above three properties.
\end{enumerate}
\end{definition}

Lexicographic ordering of codewords means codewords are ordered in an ascending manner following the rule $0 < 1 < \alpha < \dots < \alpha^{q-2}$ for any symbol, and the symbol significance reduces from left to right. In particular, starting from the left, we say $\bold{c}_{u_1} < \bold{c}_{u_2}$ if and only if for the first symbol position the two codewords differ at, $\bold{c}_{u_1}$ has a ``less'' symbol than that of $\bold{c}_{u_2}$. We omit writing ``$\forall \delta$'' inside sets for simplicity.

Let $c$ be an element in GF($q$). Define $a \triangleq \mathcal{L}(c)$ as the Flash charge level equivalent to symbol $c$, which is given by:
\begin{align}\label{eqn_gflog}
a \triangleq \mathcal{L}(c) \triangleq \left\{\begin{matrix}0, \textup{ } &c = 0,
\\ \textup{gflog}_\alpha(c)+1, \textup{ } &\textup{otherwise},
\end{matrix}\right.
\end{align}
where $\textup{gflog}_\alpha(c)$ returns the power of the GF element $c$ with $\textup{gflog}_\alpha(1)=0$. Thus, the set of charge levels equivalent to GF($q$) is $\{0, 1, 2, 3, \dots, q-1\}$, and the set of charge-level patterns equivalent to $\mathcal{Q}^q_x$ in (\ref{eqn_setq}) is:
\begin{equation}\label{eqn_setqeq}
\{(q-1)\mu(q-1), (q-1)\boldsymbol{\mu}^2(q-1), \dots,  (q-1)\boldsymbol{\mu}^x(q-1)\},
\end{equation}
where $\boldsymbol{\mu}^r \triangleq \mathcal{L}(\delta_{r-1}) \mathcal{L}(\delta_{r-2}) \dots \mathcal{L}(\delta_0)$.

Observe that the total number of elements in $\mathcal{Q}^q_x$ is:
\begin{align}\label{eqn_cardq}
\vert \mathcal{Q}^q_x \vert &= (q-1)+(q-1)^2+\dots+(q-1)^x \nonumber \\ &= \frac{(q-1)\left [(q-1)^x-1 \right ]}{q-2}.
\end{align}
Observe also that in the case of $x=1$, the set in (\ref{eqn_setq}) reduces to $\mathcal{Q}^q_1 = \{\alpha^{q-2}\delta\alpha^{q-2}\}=\{\alpha^{q-2}0\alpha^{q-2}, \alpha^{q-2}1\alpha^{q-2}, \dots, \allowbreak \alpha^{q-2}\alpha^{q-3}\alpha^{q-2}\}$ with $\vert \mathcal{Q}^q_1 \vert = q-1$ as confirmed by (\ref{eqn_cardq}). The set of level patterns equivalent to $\mathcal{Q}^q_1$ is $\{(q-1)0(q-1), (q-1)1(q-1), \dots, (q-1)(q-2)(q-1)\}$, which is the exact same set in \cite{chee_qlc} and also in \cite{veeresh_mlc} for $q=4$. It is clear that for the binary case ($q=2$), $\mathcal{Q}^2_x$ is simply $\mathcal{A}_x$.

In \cite{ahh_loco} and \cite{ahh_aloco}, we introduced tables listing all the codewords of codes with small lengths in order to illustrate ideas. For QA-LOCO codes with $q > 2$, this is no longer feasible because the number of codewords is too large. Having said that, we refer the reader to \cite[Table~I]{ahh_aloco} to check out QA-LOCO codes $\mathcal{QC}^2_{m,1}$ (or $\mathcal{AC}_{m,1}$) for $m \in \{1, 2, \dots, 5\}$.

The partition of QA-LOCO codewords into groups is essential to deriving the cardinality and later the encoding-decoding rule. We partition the codewords in $\mathcal{QC}^q_{m,x}$, $m \geq 2$, into three groups according to the symbols they start with from the left, i.e., at their left-most symbols (LMSs), as follows.\\
\textbf{Group~1:} Codewords starting with $\delta$ at their LMS.\\
\textbf{Group~2:} Codewords starting with $\alpha^{q-2}\alpha^{q-2}$ at their LMSs.\\
\textbf{Group~3:} Codewords starting with $\alpha^{q-2}\boldsymbol{\delta}^{x+1}$ at their LMSs.\footnote{In Group~3 and with $2 \leq m \leq x+1$, there exist only $(q-1)^{m-1}$ codewords, which have fewer symbols than these LMSs, in the group. The following analysis also applies for such codewords.}

Observe that given the set of forbidden patterns $\mathcal{Q}^q_{x}$ in (\ref{eqn_setq}), there are no other symbol options for a codeword $\bold{c}$ in $\mathcal{QC}^q_{m,x}$ to have at its LMSs. Now, we are ready to enumerate QA-LOCO codewords recursively.

\begin{theorem}\label{thm_card}
The cardinality (size) of a QA-LOCO code $\mathcal{QC}^q_{m,x}$, denoted by $N_q(m,x)$, is given by:
\begin{align}\label{eqn_cardgen}
N_q(m,x) &= qN_q(m-1,x) - (q-1)N_q(m-2,x) \nonumber \\ &\hspace{+1.0em}+ (q-1)^{x+1}N_q(m-x-2,x), \textup{ } m \geq 2,
\end{align}
where the defined cardinalities are:
\begin{equation}\label{eqn_cardinit}
N_q(m,x) \triangleq (q-1)^m, \textup{ } m \leq 0, \text{ and } N_q(1,x) \triangleq q.
\end{equation}
\end{theorem}

\begin{IEEEproof}
We use the group structure stated above to prove the recursive formula (\ref{eqn_cardgen}).

\textbf{Group~1:} Each codeword in Group~1 in $\mathcal{QC}^q_{m,x}$ starts with $\delta$ from the left, and therefore corresponds to a codeword in $\mathcal{QC}^q_{m-1,x}$ such that they share the $m-1$ right-most symbols (RMSs). This correspondence is surjective. Since $\delta$ is in $\{0, 1, \alpha, \dots, \alpha^{q-3}\}$, the correspondence is $q-1$ codewords of length $m$ to $1$ codeword of length $m-1$. Thus, the cardinality of Group~1 in $\mathcal{QC}^q_{m,x}$ is given by:
\begin{equation}\label{eqn_cardg1}
N_{q,1}(m,x) = (q-1)N_q(m-1,x).
\end{equation}

\textbf{Group~2:} Each codeword in Group~2 in $\mathcal{QC}^q_{m,x}$ starts with $\alpha^{q-2}\alpha^{q-2}$ from the left, and therefore corresponds to a codeword in $\mathcal{QC}^q_{m-1,x}$ that starts with $\alpha^{q-2}$ from the left such that they share the $m-2$ RMSs. This correspondence is bijective. The codewords in $\mathcal{QC}^q_{m-1,x}$ that start with $\alpha^{q-2}$ from the left are obtained by excluding the codewords in $\mathcal{QC}^q_{m-1,x}$ that start with $\delta$ from the left (the codewords of Group~1 in $\mathcal{QC}^q_{m-1,x}$) from all the codewords in $\mathcal{QC}^q_{m-1,x}$. Thus, the cardinality of Group~2 in $\mathcal{QC}^q_{m,x}$ is given by:
\begin{align}\label{eqn_cardg2}
N_{q,2}(m,x) &= N_q(m-1,x) - N_{q,1}(m-1,x) \nonumber \\ &=N_q(m-1,x) - (q-1)N_q(m-2,x),
\end{align}
where the second equality in (\ref{eqn_cardg2}) is reached aided by (\ref{eqn_cardg1}) to compute $N_{q,1}(m-1,x)$.

\textbf{Group~3:} Each codeword in Group~3 in $\mathcal{QC}^q_{m,x}$ starts with $\alpha^{q-2}\boldsymbol{\delta}^{x+1}$ from the left, and therefore corresponds to a codeword in $\mathcal{QC}^q_{m-x-1,x}$ that starts with $\delta$ from the left such that they share the $m-x-2$ RMSs. This correspondence is surjective. Since $\delta$ is in $\{0, 1, \alpha, \dots, \alpha^{q-3}\}$, the correspondence is $\Pi_{i=m-x-1}^{m-2}(q-1)=(q-1)^x$ codewords (each $\delta$ requires $\times (q-1)$) of length $m$ to $1$ codeword of length $m-x-1$. The codewords in $\mathcal{QC}^q_{m-x-1,x}$ that start with $\delta$ from the left are the codewords of Group~1 in $\mathcal{QC}^q_{m-x-1,x}$. Thus, the cardinality of Group~3 in $\mathcal{QC}^q_{m,x}$ is given by:
\begin{align}\label{eqn_cardg3}
N_{q,3}(m,x) &= (q-1)^x N_{q,1}(m-x-1,x) \nonumber \\ &= (q-1)^{x+1} N_q(m-x-2,x),
\end{align}
where the second equality in (\ref{eqn_cardg3}) is reached aided by (\ref{eqn_cardg1}) to compute $N_{q,1}(m-x-1,x)$.

Now, the cardinality of $\mathcal{QC}^q_{m,x}$ is computed as follows using (\ref{eqn_cardg1}), (\ref{eqn_cardg2}), and (\ref{eqn_cardg3}):
\vspace{-0.3em}\begin{align}
N_q(m,x) &= \sum_{\ell=1}^3 N_{q,\ell}(m,x) \nonumber \\ &= qN_q(m-1,x) - (q-1)N_q(m-2,x) \nonumber \\ &\hspace{+1.0em}+ (q-1)^{x+1}N_q(m-x-2,x), \nonumber
\end{align}
which completes the proof.
\end{IEEEproof}

Observe that substituting $q=2$ in (\ref{eqn_cardgen}) and (\ref{eqn_cardinit}) yields:
\begin{align}\label{eqn_cardq2}
N_2(m,x) &= 2N_2(m-1,x) - N_2(m-2,x) \nonumber \\ &\hspace{+1.0em}+ N_2(m-x-2,x), \textup{ } m \geq 2,
\end{align}
where the defined cardinalities are:
\begin{equation}\label{eqn_cardq2i}
N_2(m,x) \triangleq 1, \textup{ } m \leq 0, \text{ and } N_2(1,x) \triangleq 2.
\end{equation}
These are the same cardinality equations of an A-LOCO code $\mathcal{AC}_{m,x}$ (binary), which is $\mathcal{QC}^2_{m,x}$, as derived in \cite{ahh_aloco}.

\begin{example}\label{example1}
Consider the QA-LOCO codes $\mathcal{QC}^4_{m,1}$ ($q=4$ and $x=1$) with $m \in \{2, 3, \dots, 6\}$. From (\ref{eqn_cardinit}), the defined cardinalities needed here are:
\begin{equation}
N_4(-1,1) \triangleq 3^{-1}, \textup{ } N_4(0,1) \triangleq 1, \text{ and } N_4(1,1) \triangleq 4. \nonumber
\end{equation}
The cardinalities of the aforementioned QA-LOCO codes are:
\begin{align}
N_4(2,1) &= 4N_4(1,1) - 3N_4(0,1) + 9N_4(-1,1) = 16, \nonumber \\
N_4(3,1) &= 4N_4(2,1) - 3N_4(1,1) + 9N_4(0,1) = 61, \nonumber \\
N_4(4,1) &= 4N_4(3,1) - 3N_4(2,1) + 9N_4(1,1) = 232, \nonumber \\
N_4(5,1) &= 4N_4(4,1) - 3N_4(3,1) + 9N_4(2,1) = 889, \text{ and} \nonumber \\
N_4(6,1) &= 4N_4(5,1) - 3N_4(4,1) + 9N_4(3,1) = 3409. \nonumber
\end{align}
\end{example}

Theorem~\ref{thm_card} is a key result in the analysis of QA-LOCO codes. The theorem provides insights regarding how the codewords of a QA-LOCO code of a specific length relate to the codewords of QA-LOCO codes of smaller lengths. As we shall see shortly, Theorem~\ref{thm_card} and the insights it provides are fundamental to the derivation of the encoding-decoding rule, to the rate discussion, and to the algorithms.

\section{QA-LOCO Encoding-Decoding Rule}\label{sec_rule}

Now, we derive a formula that relates the lexicographic index of a QA-LOCO codeword to the codeword itself. We call this formula the encoding-decoding rule of QA-LOCO codes since it is the foundation of the QA-LOCO encoding and decoding algorithms presented in Section~\ref{sec_algr}.

We define a QA-LOCO codeword of length $m$ symbols as $\bold{c} \triangleq c_{m-1} c_{m-2} \dots c_0$ in $\mathcal{QC}^q_{m,x}$. The index of a QA-LOCO codeword $\bold{c}$ in $\mathcal{QC}^q_{m,x}$ is denoted by $g(m, x, \bold{c})$, which is sometimes abbreviated to $g(\bold{c})$ for simplicity. For each symbol $c_i$, we define its level-equivalent $a_i \triangleq \mathcal{L}(c_i)$ as shown in (\ref{eqn_gflog}), with $c_i \triangleq 0$ and $a_i \triangleq 0$ for $i \geq m$. The same notation applies for a QA-LOCO codeword of length $m+1$, $\bold{c}'$ in $\mathcal{QC}^q_{m+1,x}$, and a QA-LOCO codeword of length $m-x$, $\bold{c}''$ in $\mathcal{QC}^q_{m-x,x}$. Our lexicographic index $g(\bold{c})$ is in $\{0, 1, \dots, N_q(m,x)-1\}$.

For each codeword symbol $c_i$, define \textbf{Condition (*)} as the condition that $c_{i+k_i} \dots c_{i+2} c_{i+1} = \alpha^{q-2} \boldsymbol{\delta}^{k_i-1}$ for some $k_i \in \{1, 2, \dots, x\}$. Condition (*) can also be written as $a_{i+k_i} \dots a_{i+2} a_{i+1} = (q-1) \boldsymbol{\mu}^{k_i-1}$ for some $k_i \in \{1, 2, \dots, \allowbreak x\}$. Recall that $\boldsymbol{\mu}^r \in \{0, 1, \dots, q-2\}^r$. For example, for a QA-LOCO code with $q=4$, $m \geq 7$, and $x=3$, if we have $c_6c_5c_4c_3 = \alpha^2\alpha1\alpha$ then, $k_5 = 1$, $k_4 = 2$, and $k_3 = 3$.

The following theorem introduces the encoding-decoding rule of QA-LOCO codes. Observe that indexing is straightforward for the case of $m=1$.

\begin{theorem}\label{thm_rule}
Consider a QA-LOCO code $\mathcal{QC}^q_{m,x}$ with $m \geq 2$. Let $\bold{c}$ be a QA-LOCO codeword in $\mathcal{QC}^q_{m,x}$. The relation between the lexicographic index $g(\bold{c})$ of this codeword and the codeword itself is given by:
\begin{equation}\label{eqn_rule}
g(\bold{c}) = \sum_{i=0}^{m-1} a_i (q-1)^{\gamma_i} N_q(i-\gamma_i,x),
\end{equation}
where $\gamma_i$ for symbol $c_i$ is computed as follows:
\begin{align}\label{eqn_gammai}
\gamma_i = \left\{\begin{matrix}x-k_i+1, \textup{ } &\text{$k_i$ satisfying (*) exists},
\\ 0, \textup{ } &\text{otherwise}.
\end{matrix}\right.
\end{align}
Starting from the left (LMS), parameter $k_i \in \{1, 2, \dots, x\}$, if exists, represents the backward distance in symbols from $c_i$ to the nearest $\alpha^{q-2}$ symbol. Note that $\gamma_{m-1} = 0$.
\end{theorem}

\begin{IEEEproof}
We prove Theorem~\ref{thm_rule} by induction.

\textbf{Base:} The base case is the case of $m=2$. Using (\ref{eqn_cardgen}) and (\ref{eqn_cardinit}), the number of codewords in $\mathcal{QC}^q_{2,x}$ is:
\begin{align}\label{eqn_cardbase}
N_q(2,x) &= qN_q(1,x) - (q-1)N_q(0,x) \nonumber \\ &\hspace{+1.0em}+ (q-1)^{x+1}N_q(-x,x) \nonumber \\ &= q^2 - (q-1) + (q-1) = q^2.
\end{align}
These $q^2$ codewords are in lexicographic order: $00, 01, \allowbreak \dots, 0\alpha^{q-2}$ followed by $10, 11, \dots, 1\alpha^{q-2}$, \dots, followed by $\alpha^{q-2}0, \alpha^{q-2}1, \dots, \alpha^{q-2}\alpha^{q-2}$. We want to prove that the index obtained from (\ref{eqn_rule}) for each codeword matches its index in the aforementioned order.

First, consider the codewords in $\mathcal{QC}^q_{2,x}$ that start with $\delta$ from the left, i.e., $\bold{c} = \delta c_0$. Since $\gamma_1=\gamma_0=0$ from (\ref{eqn_gammai}), using (\ref{eqn_rule}) and (\ref{eqn_cardinit}) for such codewords gives:
\begin{align}\label{eqn_base1}
g(\bold{c}) &= a_1 N_q(1,x) + a_0 N_q(0,x) \nonumber \\ &= \mathcal{L}(\delta)q + \mathcal{L}(c_0),
\end{align}
which is indeed the correct indexing formula. For example, consider the case of $q=4$. The codeword $11$ is the $5$th in order. From (\ref{eqn_base1}), $g(\bold{c}) = \mathcal{L}(1) \times 4 + \mathcal{L}(1) = 4+1 = 5$. The codeword $\alpha \alpha^2$ is the $11$th in order. From (\ref{eqn_base1}), $g(\bold{c}) = \mathcal{L}(\alpha) \times 4 + \mathcal{L}(\alpha^2) = 8+3 = 11$.

Second, consider the codewords in $\mathcal{QC}^q_{2,x}$ that start with $\alpha^{q-2}$ from the left, i.e., $\bold{c} = \alpha^{q-2} c_0$. For $c_1$, $\gamma_1=0$ form (\ref{eqn_gammai}). For $c_0$, $k_0=1$, and therefore form (\ref{eqn_gammai}), $\gamma_0=x$. Using (\ref{eqn_rule}) and (\ref{eqn_cardinit}) for such codewords gives:
\begin{align}\label{eqn_base2}
g(\bold{c}) &= a_1 N_q(1,x) + a_0 (q-1)^x N_q(-x,x) \nonumber \\ &= \mathcal{L}(\alpha^{q-2})q + \mathcal{L}(c_0) = (q-1)q + \mathcal{L}(c_0),
\end{align}
which is indeed the correct indexing formula. For example, consider the case of $q=4$. The codeword $\alpha^2 \alpha$ is the $14$th in order. From (\ref{eqn_base2}), $g(\bold{c}) = 3 \times 4 + \mathcal{L}(\alpha) = 12+2 = 14$. Note that from (\ref{eqn_cardinit}), $N_q(1,x) \triangleq q$ and $N_q(0,x) \triangleq 1$, for all $x \in \{1, 2, \dots\}$.

\textbf{Assumption:} We assume that (\ref{eqn_rule}) is true for all the QA-LOCO codes $\mathcal{QC}^q_{\overline{m},x}$, $\overline{m} \in \{2, 3, \dots, m\}$. Mathematically, we assume the following:
\begin{equation}\label{eqn_assum1}
g(\overline{m},x,\overline{\bold{c}}) = \sum_{i=0}^{\overline{m}-1} \overline{a}_i (q-1)^{\overline{\gamma}_i} N_q(i-\overline{\gamma}_i,x),
\end{equation}
where $\overline{\bold{c}}$ is in $\mathcal{QC}^q_{\overline{m},x}$. The symbols of $\overline{\bold{c}}$ are $\overline{c}_i$, $i \in \{0, 1, \allowbreak \dots, \overline{m}-1\}$. For each $\overline{c}_i$, $\overline{a}_i \triangleq \mathcal{L}(\overline{c}_i)$ is its level-equivalent defined as in (\ref{eqn_gflog}), and $\overline{\gamma}_i$ is defined as in (\ref{eqn_gammai}).

\textbf{To be proved:} We want to prove that given the base and the assumption, (\ref{eqn_rule}) is also true for the QA-LOCO code $\mathcal{QC}^q_{m+1,x}$. In particular, we want to prove that:
\begin{equation}\label{eqn_provem}
g(m+1,x,\bold{c}') = \sum_{i=0}^{m} a'_i (q-1)^{\gamma'_i} N_q(i-\gamma'_i,x),
\end{equation}
where $\gamma'_i$ is defined for each $c'_i$ as in (\ref{eqn_gammai}), and it is a function of $x$ and $k'_i$ that depends on symbols left to $c'_i$.

We reuse our group structure to prove (\ref{eqn_provem}). We prove that (\ref{eqn_provem}) is true for the three groups in the QA-LOCO code of length $m+1$, which means it is true for the entire code. Note that our group structure can be defined for a QA-LOCO code of any length. We also reuse the codeword correspondence from the proof of Theorem~\ref{thm_card}, with $m+1$ replacing $m$.

\textbf{Group~1:} The codewords in Group~1 in $\mathcal{QC}^q_{m+1,x}$ start at index $0$, and the same applies for the corresponding codewords in $\mathcal{QC}^q_{m,x}$ (recall the lexicographic ordering rule from the start of Section~\ref{sec_card}). The correspondence here is surjective. Thus, the shift in codeword indices between $\bold{c}'$ in $\mathcal{QC}^q_{m+1,x}$ and the corresponding $\bold{c}$ in $\mathcal{QC}^q_{m,x}$ here depends on the value of $\delta$ at the LMS $c'_m$ of $\bold{c}'$. In particular,
\begin{align}\label{eqn_shift1}
g(m+1,x,\bold{c}')-g(m,x,\bold{c}) = \mathcal{L}(c'_m)N_q(m,x).
\end{align}
For example, if $c'_m=0$, the shift has to be $0$, while if $c'_m=\alpha$, the shift has to be $2N_q(m,x)$. Next, using (\ref{eqn_assum1}):
\begin{align}\label{eqn_prove1}
&g(m+1,x,\bold{c}') \nonumber \\ &= a'_m N_q(m,x) + \sum_{i=0}^{m-1} a_i (q-1)^{\gamma_i} N_q(i-\gamma_i,x).
\end{align}
Observe that $\gamma'_m = 0$, and because $c'_m \neq \alpha^{q-2}$, $\gamma'_{m-1}=0$ from (\ref{eqn_gammai}). On the other hand, $\gamma_{m-1} = 0$. Since $\bold{c}'$ and $\bold{c}$ share the $m$ RMSs and $\gamma'_{m-1}=\gamma_{m-1}$, (\ref{eqn_prove1}) can be written as:
\begin{align}\label{eqn_prove2}
g(m+1,x,\bold{c}') &= a'_m (q-1)^{\gamma'_m} N_q(m-\gamma'_m,x) \nonumber \\ &\hspace{+1.0em}+ \sum_{i=0}^{m-1} a'_i (q-1)^{\gamma'_i} N_q(i-\gamma'_i,x).
\end{align}
Consequently, we get:
\begin{equation}\label{eqn_ruleg1}
g(m+1,x,\bold{c}') = \sum_{i=0}^{m} a'_i (q-1)^{\gamma'_i} N_q(i-\gamma'_i,x).
\end{equation}

\textbf{Group~2:} The codewords in Group~2 in $\mathcal{QC}^q_{m+1,x}$ start right after Groups~1 and 3 in $\mathcal{QC}^q_{m+1,x}$, and the corresponding codewords in $\mathcal{QC}^q_{m,x}$ start right after Group~1 in $\mathcal{QC}^q_{m,x}$ (recall the lexicographic ordering rule from the start of Section~\ref{sec_card}). Moreover, the correspondence here is bijective. Thus, the shift in codeword indices between $\bold{c}'$ in $\mathcal{QC}^q_{m+1,x}$ and the corresponding $\bold{c}$ in $\mathcal{QC}^q_{m,x}$ here is:
\begin{align}\label{eqn_shift2}
&g(m+1,x,\bold{c}')-g(m,x,\bold{c}) \nonumber \\ &= N_{q,1}(m+1,x) + N_{q,3}(m+1,x) - N_{q,1}(m,x) \nonumber \\ &= (q-1)N_q(m,x) + (q-1)^{x+1}N_q(m-x-1,x) \nonumber \\ &\hspace{+1.0em}- (q-1)N_q(m-1,x),
\end{align}
where the second equality in (\ref{eqn_shift2}) is obtained aided by (\ref{eqn_cardg1}) and (\ref{eqn_cardg3}). Next, using (\ref{eqn_assum1}):
\begin{align}\label{eqn_prove3}
&g(m+1,x,\bold{c}') \nonumber \\ &= (q-1)N_q(m,x) + (q-1)^{x+1}N_q(m-x-1,x) \nonumber \\ &\hspace{+1.0em}- (q-1)N_q(m-1,x) + \sum_{i=0}^{m-1} a_i (q-1)^{\gamma_i} N_q(i-\gamma_i,x).
\end{align}

Since $c_{m-1}=\alpha^{q-2}$, which results in $a_{m-1}=q-1$, and $\gamma_{m-1}=0$, the summation term in (\ref{eqn_prove3}) can be expanded as:
\begin{align}\label{eqn_prove4}
g(m,x,\bold{c}) &= (q-1)N_q(m-1,x) \nonumber \\ &\hspace{+1.0em}+ \sum_{i=0}^{m-2} a_i (q-1)^{\gamma_i} N_q(i-\gamma_i,x).
\end{align}
Substituting (\ref{eqn_prove4}) in (\ref{eqn_prove3}) results in:
\begin{align}\label{eqn_prove5}
&g(m+1,x,\bold{c}') \nonumber \\ &= (q-1)N_q(m,x) + (q-1)^{x+1}N_q(m-x-1,x) \nonumber \\ &\hspace{+1.0em}+ \sum_{i=0}^{m-2} a_i (q-1)^{\gamma_i} N_q(i-\gamma_i,x).
\end{align}

Here, $c'_m=c'_{m-1}=\alpha^{q-2}$, which results in $a'_m=a'_{m-1}=q-1$. Observe that $\gamma'_m = 0$, and because $c'_m=\alpha^{q-2}$, $k'_{m-1}=1$, and therefore $\gamma'_{m-1}=x$ from (\ref{eqn_gammai}). Moreover, because $c'_{m-1}=c_{m-1}=\alpha^{q-2}$, $k'_{m-2}=k_{m-2}=1$, and therefore $\gamma'_{m-2}=\gamma_{m-2}=x$ from (\ref{eqn_gammai}). Since $\bold{c}'$ and $\bold{c}$ share the $m-1$ RMSs and $\gamma'_{m-2}=\gamma_{m-2}$, (\ref{eqn_prove5}) can be written as:
\begin{align}\label{eqn_prove6}
g(m+1,x,\bold{c}') &= a'_m (q-1)^{\gamma'_m} N_q(m-\gamma'_m,x) \nonumber \\ &\hspace{+1.0em}+ a'_{m-1} (q-1)^{\gamma'_{m-1}} N_q(m-1-\gamma'_{m-1},x) \nonumber \\ &\hspace{+1.0em}+ \sum_{i=0}^{m-2} a'_i (q-1)^{\gamma'_i} N_q(i-\gamma'_i,x).
\end{align}
Consequently, we get:
\begin{equation}\label{eqn_ruleg2}
g(m+1,x,\bold{c}') = \sum_{i=0}^{m} a'_i (q-1)^{\gamma'_i} N_q(i-\gamma'_i,x).
\end{equation}

\textbf{Group~3:} The codewords in Group~3 in $\mathcal{QC}^q_{m+1,x}$ start right after Group~1 in $\mathcal{QC}^q_{m+1,x}$, and the corresponding codewords in $\mathcal{QC}^q_{m-x,x}$ start at index $0$ (recall the lexicographic ordering rule from the start of Section~\ref{sec_card}). The correspondence here is surjective. Thus, the shift in codeword indices between $\bold{c}'$ in $\mathcal{QC}^q_{m+1,x}$ and the corresponding $\bold{c}''$ in $\mathcal{QC}^q_{m-x,x}$ here depends on the values in the sequence $\boldsymbol{\delta}^x$, which follows the symbol $c'_m=\alpha^{q-2}$ (the LMS), at $c'_{m-1}, \allowbreak c'_{m-2}, \dots, c'_{m-x}$ of $\bold{c}'$. At each symbol $c'_{m-j}$, $j \in \{1, 2, \dots, x\}$, an additional shift of $\mathcal{L}(c'_{m-j}) (q-1)^{x-j} N_{q,1}(m-x,x)$ should be added. Putting all terms together results in:
\begin{align}\label{eqn_shift3}
&g(m+1,x,\bold{c}')-g(m-x,x,\bold{c}'') \nonumber \\ &= N_{q,1}(m+1,x) + \sum_{j=1}^x \mathcal{L}(c'_{m-j}) (q-1)^{x-j} N_{q,1}(m-x,x).
\end{align}
Next, using (\ref{eqn_assum1}) and also (\ref{eqn_cardg1}) to compute $N_{q,1}(m+1,x)$ and $N_{q,1}(m-x,x)$, we get:
\begin{align}\label{eqn_prove7}
g(m+1,x,\bold{c}') &= (q-1) N_q(m,x) \nonumber \\ &\hspace{-1.0em}+ \hspace{+0.6em}\sum_{j=1}^x a'_{m-j} (q-1)^{x-j+1} N_q(m-x-1,x) \nonumber \\ &\hspace{-1.0em}+ \sum_{i=0}^{m-x-1} a''_i (q-1)^{\gamma''_i} N_q(i-\gamma''_i,x).
\end{align}

We keep our focus on the symbols $c'_{m-j}$, for all $j \in \{1, \allowbreak 2, \dots, x\}$. Consider a specific $c'_{m-j}$. Since $c'_m=\alpha^{q-2}$, and until $c'_{m-x}$ (from the LMS $c'_m$ going right) it is guaranteed that there are no other $\alpha^{q-2}$ symbols, $k'_{m-j}=j$. Thus, $\gamma'_{m-j}=x-j+1$ from (\ref{eqn_gammai}). Moreover, we can write the term $m-x-1$ as $m-j-(x-j+1)=m-j-\gamma'_{m-j}$, for all $j \in \{1, 2, \dots, x\}$. Consequently, we get:
\begin{align}\label{eqn_prove8}
&\sum_{j=1}^x a'_{m-j} (q-1)^{x-j+1} N_q(m-x-1,x) \nonumber \\ &= \hspace{+0.6em}\sum_{j=1}^x a'_{m-j} (q-1)^{\gamma'_{m-j}} N_q(m-j-\gamma'_{m-j},x) \nonumber \\ &= \sum_{i=m-x}^{m-1} a'_i (q-1)^{\gamma'_i} N_q(i-\gamma'_i,x).
\end{align}
The last equality in (\ref{eqn_prove8}) is reached using the simple transformation of variables $i=m-j$.

Here, $c'_m=\alpha^{q-2}$, which results in $a'_m=q-1$. Observe that $\gamma'_m = 0$, and that (\ref{eqn_prove8}) covers all the symbols $c'_{m-j}$, for all $j \in \{1, 2, \dots, x\}$. Moreover, because there does not exist $k'_{m-x-1}$ in $\{1, 2, \dots, x\}$ that satisfies Condition (*) for $c'_{m-x-1}$, $\gamma'_{m-x-1}=0$ from (\ref{eqn_gammai}). It is also the case that $\gamma''_{m-x-1}=0$. Since $\bold{c}'$ and $\bold{c}''$ share the $m-x-1$ RMSs, $c'_{m-x-1}=c''_{m-x-1}$, and $\gamma'_{m-x-1}=\gamma''_{m-x-1}$, (\ref{eqn_prove7}) can be written, aided by (\ref{eqn_prove8}), as:
\begin{align}\label{eqn_prove9}
g(m+1,x,\bold{c}') &= a'_m (q-1)^{\gamma'_m} N_q(m-\gamma'_m,x) \nonumber \\ &\hspace{-1.0em}+ \sum_{i=m-x}^{m-1} a'_i (q-1)^{\gamma'_i} N_q(i-\gamma'_i,x) \nonumber \\ &\hspace{-1.0em}+ \sum_{i=0}^{m-x-1} a'_i (q-1)^{\gamma'_i} N_q(i-\gamma'_i,x).
\end{align}
Consequently, we get:
\begin{equation}\label{eqn_ruleg3}
g(m+1,x,\bold{c}') = \sum_{i=0}^{m} a'_i (q-1)^{\gamma'_i} N_q(i-\gamma'_i,x).
\end{equation}

From (\ref{eqn_ruleg1}), (\ref{eqn_ruleg2}), and (\ref{eqn_ruleg3}), (\ref{eqn_provem}) is proved for all three groups in $\mathcal{QC}^q_{m+1,x}$, which means (\ref{eqn_provem}) is proved for the entire code. This completes the proof by induction, and thus, the encoding-decoding rule in (\ref{eqn_rule}) is proved for any QA-LOCO code $\mathcal{QC}^q_{m,x}$ with $q \geq 2$, $m \geq 2$, and $x \geq 1$.
\end{IEEEproof}

Observe that substituting $q=2$ in (\ref{eqn_rule}) yields:
\begin{equation}\label{eqn_ruleq2}
g(\bold{c}) = \sum_{i=0}^{m-1} a_i N_2(i-\gamma_i,x),
\end{equation}
where for $c_i \neq 0$, i.e., $a_i \neq 0$, $\gamma_i$ here is either $x$ in the case of $c_{i+1}=1$ or $0$ in the case of $c_{i+1}=0$. Thus, $\gamma_i$ can be written as $\mathcal{L}(c_{i+1})x=a_{i+1}x$. Substituting $\gamma_i=a_{i+1}x$ in (\ref{eqn_ruleq2}) gives the rule of an A-LOCO code $\mathcal{AC}_{m,x}$ (binary), which is $\mathcal{QC}^2_{m,x}$, as derived in \cite{ahh_aloco}.

\begin{example}\label{example2}
We use (\ref{eqn_rule}) to compute the index of two QA-LOCO codewords in $\mathcal{QC}^4_{6,2}$ ($q=4$, $m=6$, and $x=2$). Using Theorem~\ref{thm_card}, the required cardinalities are $N_4(-1,2) \triangleq 3^{-1}$, $N_4(0,2) \triangleq 1$, $N_4(1,2) \triangleq 4$, $N_4(2,2) = 16$, $N_4(3,2) = 61$, $N_4(4,2) = 223$, and $N_4(5,2) = 817$.

The first codeword is the $334$th codeword $011\alpha^20\alpha$. This codeword has $a_5=0$, $a_4=a_3=1$, $a_2=3$, $a_1=0$, and $a_0=2$. From (\ref{eqn_gammai}), we get $\gamma_5=\gamma_4=\gamma_3=\gamma_2=0$, $\gamma_1=x=2$, and $\gamma_0=x-1=1$. Thus, from (\ref{eqn_rule}):
\begin{align}
g(\bold{c}) &= \sum_{i=0}^{5} a_i (3^{\gamma_i}) N_4(i-\gamma_i,2) \nonumber \\ &= N_4(4,2) + N_4(3,2) + 3N_4(2,2) + 6N_4(-1,2) \nonumber \\ &= 223 + 61 + 3 \times 16 + 6 \times 3^{-1} = 334, \nonumber
\end{align}
which is the correct index.

The second codeword is the $1850$th codeword $\alpha0\alpha^2\alpha^2\alpha0$. This codeword has $a_5=2$, $a_4=0$, $a_3=a_2=3$, $a_1=2$, and $a_0=0$. From (\ref{eqn_gammai}), we get $\gamma_5=\gamma_4=\gamma_3=0$, $\gamma_2=\gamma_1=x=2$, and $\gamma_0=x-1=1$. Thus, from (\ref{eqn_rule}):
\begin{align}
g(\bold{c}) &= \sum_{i=0}^{5} a_i (3^{\gamma_i}) N_4(i-\gamma_i,2) \nonumber \\ &= 2N_4(5,2) + 3N_4(3,2) + 27N_4(0,2) + 18N_4(-1,2) \nonumber \\ &= 2 \times 817 + 3 \times 61 + 27 \times 1 + 18 \times 3^{-1} = 1850, \nonumber
\end{align}
which is the correct index.
\end{example}

Theorem~\ref{thm_rule} is the key result behind the simple, reconfigurable QA-LOCO encoding and decoding we offer. The theorem provides one-to-one mapping from an index to the corresponding codeword, which is the encoding, and one-to-one demapping from a codeword to the corresponding index, which is the decoding. Section~\ref{sec_algr} provides algorithms for QA-LOCO encoding and decoding, as well as a discussion of their reconfigurability.

\section{Achievable Rates and Comparisons}\label{sec_rate}

Before we introduce the achievable rates of QA-LOCO codes and make comparisons with other codes, we first discuss how to achieve bridging and self-clocking.

Bridging is required in order to prevent forbidden patterns from appearing while transitioning from a codeword into the next one \cite{ahh_loco}. Consider the QA-LOCO code $\mathcal{QC}^4_{5,1}$ ($q=4$, $m=5$, and $x=1$). Assume that we are about to write the following two consecutive codewords on an MLC ($4$ levels per cell) Flash device: $01\alpha\alpha^2\alpha^2$ and $1\alpha^2001$. The stream containing the two consecutive codewords to be written on ten consecutive cells is $01\alpha\alpha^2\underline{\alpha^21\alpha^2}001$, and it does contain the forbidden pattern $\alpha^21\alpha^2$. Bridging fixes such a problem.

Let $e \triangleq \alpha^{q-2}$. We perform bridging in a QA-LOCO code $\mathcal{QC}^q_{m,x}$ via adding bridging patterns as follows:
\begin{enumerate}
\item If the RMS of a codeword and the LMS of the next codeword are both $\alpha^{q-2}$'s, bridge with $\bold{e}^x$, i.e., bridge with $x$ consecutive $e \triangleq \alpha^{q-2}$ symbols ($x$ consecutive cells programmed to level $q-1$).
\item Otherwise, bridge with $\bold{0}^x$, i.e., bridge with $x$ consecutive $0$ symbols ($x$ consecutive unprogrammed cells).
\end{enumerate}

Applying this bridging method to the above scenario results in the following stream $01\alpha\alpha^2\underline{\alpha^201\alpha^2}001$. Bridging with $0$ between the two codewords prevents the forbidden pattern from appearing across the codewords.

Our bridging is not only simple, but also optimal in the sense that it provides the maximum protection from ICI for the symbols at the edges of QA-LOCO codewords. Note also that this bridging helps us reduce the number of codewords to be removed from the QA-LOCO code such that we achieve self-clocking to only two codewords as we discuss below.\footnote{With more advanced bridging for $q > 2$, this number can be reduced to one codeword to be removed such that we achieve self-clocking. However, the reduction from two to one practically has no effect on the rate.}

Self-clocking is required in order to maintain calibration~of the system \cite{immink_surv, ahh_aloco}. Self-clocked constrained codes do not allow long streams of the same symbol to be written (transmitted). Given our bridging method illustrated above for a QA-LOCO code $\mathcal{QC}^q_{m,x}$, even if we repeat a same-symbol codeword consecutive times in a stream, as long as this symbol is in GF($q$)$\setminus\{0, \alpha^{q-2}\}$, bridging will guarantee that two transitions to then from a different symbol ($0$) occur right before each new codeword in the stream. This does not happen with only two same-symbol codewords, which are $\bold{0}^m$ and $\bold{e}^m$, $e \triangleq \alpha^{q-2}$. Consequently, these are the only codewords we need to remove from $\mathcal{QC}^q_{m,x}$ to achieve self-clocking.

\begin{definition}\label{def_cqaloco}
Let $\mathcal{QC}^q_{m,x}$ be a QA-LOCO code with $q \geq 2$, $m \geq 1$, and $x \geq 1$. A self-clocked QA-LOCO code (CQA-LOCO code) $\mathcal{QC}^{q,\textup{c}}_{m,x}$ is obtained from $\mathcal{QC}^q_{m,x}$ as follows:
\begin{equation}\label{eqn_cqaloco}
\mathcal{QC}^{q,\textup{c}}_{m,x} \triangleq \mathcal{QC}^q_{m,x}\setminus\{\bold{0}^m, \bold{e}^m\}, \textup{ } e \triangleq \alpha^{q-2}.
\end{equation}
Therefore, the cardinality of the CQA-LOCO code is:
\begin{equation}\label{eqn_cqacard}
N^{\textup{c}}_q(m,x) = N_q(m,x)-2.
\end{equation}
\end{definition}

Define $k^{\textup{c}}_{\textup{eff}}$ as the maximum number of consecutive cells between two consecutive transitions (all programmed to the same level or all unprogrammed) after a stream of CQA-LOCO codewords separated by bridging patterns is written; one symbol per cell. Thus, $k^{\textup{c}}_{\textup{eff}}$ is the length of the longest run of consecutive $0$'s, $1$'s, $\alpha$'s, \dots, or $\alpha^{q-2}$'s in a stream of CQA-LOCO codewords separated by bridging patterns. The following is one scenario under which $k^{\textup{c}}_{\textup{eff}}$ is achieved:
\begin{equation}
\delta\bold{e}^{m-1} - \bold{e}^x - \bold{e}^{m-1}\delta. \nonumber
\end{equation}
As a result, $k^{\textup{c}}_{\textup{eff}}$ is given by:
\begin{equation}\label{eqn_keff}
k^{\textup{c}}_{\textup{eff}} = 2(m-1)+x,
\end{equation}
which is the same equation satisfied by LOCO codes \cite{ahh_loco} and A-LOCO codes \cite{ahh_aloco}.

\begin{table*}
\caption{Rates and Normalized Rates of Various CQA-LOCO Codes $\mathcal{QC}^{q,\textup{c}}_{m,1}$ with $q \in \{4, 8, 16, 32\}$ (for M/T/Q/P-LC Flash) and $x=1$.}
\vspace{-0.1em}
\centering
\scalebox{0.95}
{
\begin{tabular}{|c|c|c|c|c|c|c|c|c|c|c|c|}
\hline
\multicolumn{3}{|c|}{\makecell{$q=4$}} & \multicolumn{3}{|c|}{\makecell{$q=8$}} & \multicolumn{3}{|c|}{\makecell{$q=16$}} & \multicolumn{3}{|c|}{\makecell{$q=32$}} \\
\hline
{$m$} & $R^{\textup{c}}_{\textup{QA-LOCO}}$ & $R^{\textup{c},\textup{n}}_{\textup{QA-LOCO}}$ & {$m$} & $R^{\textup{c}}_{\textup{QA-LOCO}}$ & $R^{\textup{c},\textup{n}}_{\textup{QA-LOCO}}$ & {$m$} & $R^{\textup{c}}_{\textup{QA-LOCO}}$ & $R^{\textup{c},\textup{n}}_{\textup{QA-LOCO}}$ & {$m$} & $R^{\textup{c}}_{\textup{QA-LOCO}}$ & $R^{\textup{c},\textup{n}}_{\textup{QA-LOCO}}$ \\
\hline
$14$ & $1.8000$ & $0.9000$ & $18$ & $2.7895$ & $0.9298$ & $18$ & $3.7368$ & $0.9342$ & $19$ & $4.7000$ & $0.9400$ \\
\hline
$26$ & $1.8519$ & $0.9260$ & $26$ & $2.8519$ & $0.9506$ & $27$ & $3.8214$ & $0.9554$ & $29$ & $4.8000$ & $0.9600$ \\
\hline
$49$ & $1.9000$ & $0.9500$ & $44$ & $2.9111$ & $0.9704$ & $45$ & $3.8913$ & $0.9728$ & $49$ & $4.8800$ & $0.9760$ \\
\hline
$77$ & $1.9103$ & $0.9552$ & $71$ & $2.9306$ & $0.9769$ & $66$ & $3.9254$ & $0.9813$ & $70$ & $4.9155$ & $0.9831$ \\
\hline
$97$ & $1.9184$ & $0.9592$ & $103$ & $2.9519$ & $0.9840$ & $111$ & $3.9554$ & $0.9888$ & $117$ & $4.9492$ & $0.9898$ \\
\hline
Capacity & $1.9374$ & $0.9687$ & Capacity & $2.9817$ & $0.9939$ & Capacity & $3.9950$ & $0.9987$ & Capacity & $4.9987$ & $0.9997$ \\
\hline
\end{tabular}}
\label{table1}
\end{table*}

\begin{table*}
\caption{Rates and Normalized Rates of Various CQA-LOCO Codes $\mathcal{QC}^{q,\textup{c}}_{m,2}$ with $q \in \{4, 8, 16, 32\}$ (for M/T/Q/P-LC Flash) and $x=2$.}
\vspace{-0.1em}
\centering
\scalebox{0.95}
{
\begin{tabular}{|c|c|c|c|c|c|c|c|c|c|c|c|}
\hline
\multicolumn{3}{|c|}{\makecell{$q=4$}} & \multicolumn{3}{|c|}{\makecell{$q=8$}} & \multicolumn{3}{|c|}{\makecell{$q=16$}} & \multicolumn{3}{|c|}{\makecell{$q=32$}} \\
\hline
{$m$} & $R^{\textup{c}}_{\textup{QA-LOCO}}$ & $R^{\textup{c},\textup{n}}_{\textup{QA-LOCO}}$ & {$m$} & $R^{\textup{c}}_{\textup{QA-LOCO}}$ & $R^{\textup{c},\textup{n}}_{\textup{QA-LOCO}}$ & {$m$} & $R^{\textup{c}}_{\textup{QA-LOCO}}$ & $R^{\textup{c},\textup{n}}_{\textup{QA-LOCO}}$ & {$m$} & $R^{\textup{c}}_{\textup{QA-LOCO}}$ & $R^{\textup{c},\textup{n}}_{\textup{QA-LOCO}}$ \\
\hline
$20$ & $1.7273$ & $0.8636$ & $22$ & $2.7083$ & $0.9028$ & $24$ & $3.6538$ & $0.9135$ & $25$ & $4.5926$ & $0.9185$ \\
\hline
$38$ & $1.8000$ & $0.9000$ & $32$ & $2.7941$ & $0.9314$ & $34$ & $3.7500$ & $0.9375$ & $36$ & $4.7105$ & $0.9421$ \\
\hline
$57$ & $1.8305$ & $0.9153$ & $52$ & $2.8519$ & $0.9506$ & $51$ & $3.8302$ & $0.9575$ & $56$ & $4.8103$ & $0.9621$ \\
\hline
$76$ & $1.8462$ & $0.9231$ & $73$ & $2.8800$ & $0.9600$ & $73$ & $3.8800$ & $0.9700$ & $77$ & $4.8608$ & $0.9722$\\
\hline
$96$ & $1.8571$ & $0.9285$ & $108$ & $2.9091$ & $0.9697$ & $100$ & $3.9118$ & $0.9779$ & $108$ & $4.9000$ & $0.9800$ \\
\hline
Capacity & $1.8947$ & $0.9473$ & Capacity & $2.9675$ & $0.9892$ & Capacity & $3.9906$ & $0.9977$ & Capacity & $4.9975$ & $0.9995$ \\
\hline
\end{tabular}}
\label{table2}
\end{table*}

Now, we are ready to discuss the achievable rates of QA-LOCO codes. Consider a CQA-LOCO code $\mathcal{QC}^{q,\textup{c}}_{m,x}$ with cardinality $N^{\textup{c}}_q(m,x)$, which is given in (\ref{eqn_cqacard}). The length, in bits, of the messages $\mathcal{QC}^{q,\textup{c}}_{m,x}$ encodes is:
\begin{equation}
s^{\textup{c}} = \left \lfloor \log_2 N^{\textup{c}}_q(m,x) \right \rfloor = \left \lfloor \log_2 \left( N_q(m,x)-2 \right) \right \rfloor.
\end{equation}
The input information message is intentionally selected to be a binary message in order to minimize the number of omitted codewords from $\mathcal{QC}^{q,\textup{c}}_{m,x}$, and therefore maximize the rate for $q > 2$. We will give an example on that shortly. The rate of the CQA-LOCO code $\mathcal{QC}^{q,\textup{c}}_{m,x}$ then is:
\begin{equation}\label{eqn_rate}
R^{\textup{c}}_{\textup{QA-LOCO}} = \frac{s^{\textup{c}}}{m+x} = \frac{\left \lfloor \log_2 \left( N_q(m,x)-2 \right) \right \rfloor}{m+x},
\end{equation}
where $R^{\textup{c}}_{\textup{QA-LOCO}}$ is measured in information bits per coded symbol. We can normalize this rate as follows:
\begin{equation}\label{eqn_ratenorm}
R^{\textup{c},\textup{n}}_{\textup{QA-LOCO}} = \frac{\left \lfloor \log_2 \left( N_q(m,x)-2 \right) \right \rfloor}{(m+x)\log_2 q}.
\end{equation}

\begin{example}\label{example3}
Consider the CQA-LOCO code $\mathcal{QC}^{4,\textup{c}}_{9,1}$ ($q=4$, $m=9$, and $x=1$). From the recursion in Theorem~\ref{thm_card}, we can reach that $N_4(9,1) = 191518$. From (\ref{eqn_rate}), we get a rate of:
\begin{equation}
R^{\textup{c}}_{\textup{QA-LOCO}} = \frac{\left \lfloor \log_2 \left( 191518-2 \right) \right \rfloor}{9+1} = 1.7 \nonumber
\end{equation}
information bits per coded symbol. From (\ref{eqn_ratenorm}), the normalized rate is $1.7/\log_2 4 = 0.85$.

Now, suppose that we want to encode non-binary messages, with their symbols defined over GF($4$) here. The rate in this case becomes:
\begin{equation}
\overline{R}^{\textup{c}}_{\textup{QA-LOCO}} = \frac{\left \lfloor \log_4 \left( 191518-2 \right) \right \rfloor}{9+1} = 0.8. \nonumber
\end{equation}
Clearly, this is a significant rate loss compared with the $0.85$ normalized rate achieved by encoding binary information messages.\footnote{CQA-LOCO code rates that are a lot closer to the capacity of a $\mathcal{Q}^4_1$-constrained code are going to be presented in this section.} The reason is the higher number of omitted codewords when messages are non-binary. In particular, the number of omitted codewords when messages are binary here is $191516-2^{17}=60444$. This number becomes $191516-4^{8}=125980$ when messages are non-binary.
\end{example}

Except only the two codewords $\bold{0}^m$ and $\bold{e}^m$, $e \triangleq \alpha^{q-2}$, all the codewords satisfying the $\mathcal{Q}^q_x$ constraint are in the CQA-LOCO code $\mathcal{QC}^{q,\textup{c}}_{m,x}$. Additionally, the number of symbols we add for bridging is constant, which is $x$. Thus, CQA-LOCO codes are \textbf{capacity-achieving} codes, i.e., the asymptotic rate of a CQA-LOCO code matches the capacity.

Tables~\ref{table1} and \ref{table2} present the rates and the normalized rates of~CQA-LOCO codes $\mathcal{QC}^{q,\textup{c}}_{m,x}$ with $q \in \{4, 8, 16, 32\}$, various values of $m$, and $x \in \{1, 2\}$. The capacities are given in the last row of each table. We compute the capacity of a $\mathcal{Q}^q_x$-constrained code from the finite-state transition diagram (FSTD) representing the infinitude of a sequence satisfying this $\mathcal{Q}^q_x$ constraint; the capacity, in information bits per coded symbol, is the base-$2$ logarithm of the largest positive eigenvalue of the adjacency matrix corresponding to the FSTD.

Table~\ref{table1} demonstrates that for all values of $q$, the rates of CQA-LOCO codes with $x=1$ and moderate lengths reach within only $1\%$ from capacity; see the rates in the row right before the capacity row. Furthermore, Table~\ref{table2} demonstrates that for all values of $q$, the rates of CQA-LOCO codes with $x=2$ and moderate lengths reach within only $2\%$ from capacity; see the rates in the row right before the capacity row. Most important, the tables show that CQA-LOCO codes for all values of $q$ and $x$ achieve normalized rates $> 0.95$, i.e., rates $> 0.95 \log_2 q$ information bits per coded symbol, with only one exception, which is the case of $q=4$ and $x=2$. In other words, significant ICI mitigation in the Flash device can be achieved with only $5\%$ or less redundancy, even late in the lifetime of the device when $x$ can be raised to $2$.

The two tables also show the effect of increasing $q$ on the achievable rates. As $q$ increases, the sufficient rate to protect the Flash device increases. Consider QLC ($q=16$) and PLC ($q=32$) Flash devices. For $x=1$, Table~\ref{table1} shows that only about $1.9\%$ (resp., $1.7\%$) redundancy is enough at length $66$ symbols (resp., $70$ symbols) for QLC devices (resp., PLC devices). For $x=2$, Table~\ref{table2} shows that only about $3\%$ (resp., $2.8\%$) redundancy is enough at length $73$ symbols (resp., $77$ symbols) for QLC devices (resp., PLC devices). Essentially, this is telling that the ICI mitigation via CQA-LOCO codes is coming \textbf{almost for free with respect to redundancy}. Having said that, increasing $q$ results in an increase in the storage and complexity as we shall see next section.

Next, we present brief comparisons between QA-LOCO codes and other codes designed for similar goals:
\begin{enumerate}
\item It is already not easy to design FSM-based binary constrained codes with rates close to capacity \cite{ahh_loco, siegel_mr}. This task becomes even more complicated in the non-binary domain. Our QA-LOCO codes offer simple encoding and decoding because of their rule, even with $q > 2$.

\item The authors of \cite{tang_bahl} introduced $q$-ary lexicographically-ordered RLL (Q-LO-RLL) codes. However, their constraints impose a minimum number of zeros between each two consecutive non-zero symbols. This results in a significant rate loss, that is not needed, if applied for Flash. In the binary case, LOCO codes were shown in \cite{ahh_loco} to offer a better rate-complexity trade-off compared with LO-RLL codes designed for the same purpose.

\item The authors of \cite{chee_qlc} introduced enumerative $q$-ary $\mathcal{Q}^q_1$-constrained codes for Flash. While their codes are capacity-achieving and efficient, QA-LOCO codes offer simpler encoding and decoding compared with their unrank-rank approach. Additionally, the codes in \cite{chee_qlc} are only for the case of $x=1$, which means QA-LOCO codes address more general constraints.

\item We suggest that non-binary constrained codes are significantly more efficient, rate-wise, compared with binary codes. From \cite{ahh_aloco}, the capacity of a binary $\mathcal{A}_1$-constrained code ($x=1$) is $0.8114$. From Table~\ref{table2}, we can see that even for $q=4$, a self-clocked QA-LOCO code of length only $20$ symbols achieves about $6.4\%$ rate advantage with respect to the aforementioned binary capacity, and at $x=2$ (more ICI mitigation).
\end{enumerate}

\begin{remark}
A balanced binary constrained code associated with level-based (NRZ) signaling has the property that the absolute difference between the number of $1$'s and $0$'s in any stream of its codewords is bounded. Symmetric LOCO codes can be easily balanced with a minimal rate loss as shown in \cite{ahh_loco}. In the context of $q$-ary constrained codes for Flash, balancing was introduced in \cite{qin_flash} as the property that each codeword has uniform distribution for the number of instances of each symbol. Almost-balanced QA-LOCO codes can be designed with less restrictions. 
\end{remark}

\section{Algorithms and Reconfigurability}\label{sec_algr}

Now, we introduce the encoding and decoding algorithms of QA-LOCO codes, which are based on their encoding-decoding rule (\ref{eqn_rule}) of Theorem~\ref{thm_rule}. The algorithms perform the mapping-demapping between an index and the associated codeword, and thus, they are essential for enumerative techniques to offer simplicity. See \cite{laroia_const} for a conceptually connected work in the context of multi-dimensional constellations.

Algorithm~\ref{alg_enc} is the encoding algorithm of our codes. While generating a specific codeword $\bold{c}$ in the algorithm, the RMS of the previous codeword is defined as $\zeta_0$. Example~\ref{example4} illustrates how Algorithm~\ref{alg_enc} works.

\begin{algorithm}
\caption{Encoding CQA-LOCO Codes}
\begin{algorithmic}[1]
\State \textbf{Input:} Incoming stream of binary messages.
\State Set $q = \log_2$(number of levels per Flash cell).
\State Decide the value of $x$ based on system requirements.
\State Use (\ref{eqn_cardgen}) and (\ref{eqn_cardinit}) to compute $N_q(i, x)$, $i \in \{2, 3, \dots\}$.
\State Specify $m$, the smallest $i$ in Step~4 to achieve the desired rate. Then, $s^{\textup{c}} = \left \lfloor \log_2 \left( N_q(m, x) - 2 \right )  \right \rfloor$.
\State \textbf{for} each incoming message $\bold{b}$ of length $s^{\textup{c}}$ \textbf{do}
\State \hspace{2ex} Compute $g(\bold{c})=\textup{decimal}(\bold{b})+1$.
\State \hspace{2ex} Initialize $\textup{residual}$ with $g(\bold{c})$ and $c_i$ with $0$ for $i \geq m$.
\State \hspace{2ex} Initialize $\gamma_i$ with $0$ for $i \in \{0, 1, \dots, m-1\}$.
\State \hspace{2ex} \textbf{for} $i \in \{m-1, m-2, \dots, 0\}$ \textbf{do} \textit{(in order)}
\State \hspace{4ex} \textbf{for} $k_i \in \{1, 2, \dots, x\}$ \textbf{do}
\State \hspace{6ex} \textbf{if} $c_{i+k_i} = \alpha^{q-2}$ \textbf{then}
\State \hspace{8ex} Set $\gamma_i = x-k_i+1$.
\State \hspace{8ex} \textbf{break}. \textit{(exit current loop)}
\State \hspace{6ex} \textbf{end if}
\State \hspace{4ex} \textbf{end for}
\State \hspace{4ex} Set $\textup{index} = i - \gamma_i$.
\State \hspace{4ex} \textbf{if} $\textup{residual} < (q-1)^{\gamma_i} N_q(\textup{index}, x)$ \textbf{then}
\State \hspace{6ex} Encode $c_i = 0$. \textit{(level $a_i=0$)}
\State \hspace{4ex} \textbf{else if} $\textup{residual} \geq (q-1)^{\gamma_i+1} N_q(\textup{index}, x)$ \textbf{then}
\State \hspace{6ex} Encode $c_i = \alpha^{q-2}$. \textit{(level $a_i=q-1$)}
\State \hspace{6ex} $\textup{residual} \leftarrow \textup{residual} - (q-1)^{\gamma_i+1} N_q(\textup{index}, x)$.
\State \hspace{4ex} \textbf{else}
\State \hspace{6ex} \textbf{for} $a_i \in \{1, 2, \dots, q-2\}$ \textbf{do}
\State \hspace{8ex} \textbf{if} $a_i (q-1)^{\gamma_i} N_q(\textup{index}, x) \leq \textup{residual} < (a_i+1) (q-1)^{\gamma_i} N_q(\textup{index}, x)$ \textbf{then}
\State \hspace{10ex} Encode $c_i = \mathcal{L}^{-1}(a_i)$. \textit{(level $a_i = \mathcal{L}(c_i)$)}
\State \hspace{10ex} $\textup{residual} \leftarrow \textup{residual} - a_i (q-1)^{\gamma_i} N_q(\textup{index}, x)$.
\State \hspace{10ex} \textbf{break}. \textit{(exit current loop)}
\State \hspace{8ex} \textbf{end if}
\State \hspace{6ex} \textbf{end for}
\State \hspace{4ex} \textbf{end if}
\State \hspace{4ex} \textbf{if} (not first codeword) $\land$ ($i = m-1$) \textbf{then}
\State \hspace{6ex} \textbf{if} ($\zeta_0 = \alpha^{q-2}$) $\land$ ($c_{m-1} = \alpha^{q-2}$) \textbf{then}
\State \hspace{8ex} Bridge with $x$ $\alpha^{q-2}$'s, i.e., $\bold{e}^x$, before $c_{m-1}$.
\State \hspace{6ex} \textbf{else}
\State \hspace{8ex} Bridge with $x$ $0$'s, i.e., $\bold{0}^x$, before $c_{m-1}$.
\State \hspace{6ex} \textbf{end if}
\State \hspace{4ex} \textbf{end if}
\State \hspace{2ex} \textbf{end for}
\State \textbf{end for}
\State \textbf{Output:} Outgoing stream of $q$-ary CQA-LOCO codewords. \textit{(to be written on the Flash device)}
\end{algorithmic}
\label{alg_enc}
\end{algorithm}

\begin{example}\label{example4}
Consider the CQA-LOCO code $\mathcal{QC}^{4,\textup{c}}_{6,1}$ ($q=4$, $m=6$, and $x=1$). From Theorem~\ref{thm_card}, $N_4(-1,1) \triangleq 3^{-1}$, $N_4(0,1) \triangleq 1$, $N_4(1,1) \triangleq 4$, $N_4(2,1) = 16$, $N_4(3,1) = 61$, $N_4(4,1) = 232$, $N_4(5,1) = 889$, and $N_4(6,1) = 3409$. Thus, $s^{\textup{c}} = \lfloor \log_2 3407 \rfloor = 11$ bits. Now, suppose we want to encode the binary message $\bold{b}=11011001111$ via $\mathcal{QC}^{4,\textup{c}}_{6,1}$ using Algorithm~\ref{alg_enc}. From Step~7, $g(\bold{c}) = \textup{decimal}(\bold{b})+1 = 1743$, which is the initial $\textup{residual}$ from Step~8. The encoding is performed as follows (the loop in Steps~10--39):
\begin{enumerate}
\item For $i=5$, $c_6 \triangleq 0$. Thus, $\gamma_5$ stays $0$ (see Steps~11--16), and from Step~17, $\textup{index}=i=5$. Neither the condition at Step~18 nor the one at Step~20 is satisfied. Thus, the loop starting at Step~24 is entered. Since $N_4(5,1) = 889 < \textup{residual} < 2N_4(5,1) = 1778$, $c_5$ is encoded as $\mathcal{L}^{-1}(1) = 1$ from Step~26, and $\textup{residual}$ becomes $1743-889 = 854$ from Step~27.

\item For $i=4$, $c_5=1$. Thus, $\gamma_4$ stays $0$ (see Steps~11--16), and from Step~17, $\textup{index}=i=4$. The condition at Step~20 is satisfied since $\textup{residual} > 3N_4(4,1) = 696$. Thus, $c_4$ is encoded as $\alpha^2$ from Step~21, and $\textup{residual}$ becomes $854-696 = 158$ from Step~22.

\item For $i=3$, $c_4=\alpha^2$. Thus, from Steps~12 and 13, $k_3=1$ and $\gamma_3=1-1+1=1$, and from Step~17, $\textup{index}=i-1=2$. The condition at Step~20 is again satisfied since~$\textup{residual} > 9N_4(2,1) = 144$. Thus, $c_3$ is encoded as $\alpha^2$ from Step~21, and $\textup{residual}$ becomes $158-144 = 14$ from Step~22.

\item For $i=2$, $c_3=\alpha^2$. Thus, from Steps~12 and 13, $k_2=1$ and $\gamma_2=1-1+1=1$, and from Step~17, $\textup{index}=i-1=1$. Neither the condition at Step~18 nor the one at Step~20 is satisfied. Thus, the loop starting at Step~24 is entered. Since $3N_4(1,1) = 12 < \textup{residual} < 6N_4(1,1) = 24$, $c_2$ is encoded as $\mathcal{L}^{-1}(1) = 1$ from Step~26, and $\textup{residual}$ becomes $14-12 = 2$ from Step~27.

\item For $i=1$, $c_2=1$. Thus, $\gamma_1$ stays $0$ (see Steps~11--16), and from Step~17, $\textup{index}=i=1$. The condition at Step~18 is satisfied since $\textup{residual} < N_4(1,1) = 4$. Thus, $c_1$ is encoded as $0$ from Step~19, and $\textup{residual}$ stays $2$.

\item For $i=0$, $c_1=0$. Thus, $\gamma_0$ stays $0$ (see Steps~11--16), and from Step~17, $\textup{index}=i=0$. Neither the condition at Step~18 nor the one at Step~20 is satisfied. Thus, the loop starting at Step~24 is entered. Since $2N_4(0,1) = 2 = \textup{residual} < 3N_4(0,1) = 3$, $c_0$ is encoded as $\mathcal{L}^{-1}(2) = \alpha$ from Step~26, and $\textup{residual}$ becomes $2-2 = 0$ from Step~27.
\end{enumerate}
The generated codeword is then $\bold{c} = 1\alpha^2\alpha^210\alpha$, which is indeed the correct codeword. Bridging is then performed in Steps~32--38.
\end{example}

Algorithm~\ref{alg_dec} is the decoding algorithm of our codes, and it is a direct implementation of (\ref{eqn_rule}). Thus, Example~\ref{example2} illustrates how Algorithm~\ref{alg_dec} works.

\begin{algorithm}
\caption{Decoding CQA-LOCO Codes}
\begin{algorithmic}[1]
\State \textbf{Inputs:} Incoming stream of $q$-ary CQA-LOCO codewords, in addition to $q$, $m$, $x$, and $s^{\textup{c}}$.
\State Use (\ref{eqn_cardgen}) and (\ref{eqn_cardinit}) to compute $N_q(i, x)$, $i \in \{2, 3, \dots, m\}$.
\State \textbf{for} each incoming codeword $\bold{c}$ of length $m$ \textbf{do}
\State \hspace{2ex} Initialize $g(\bold{c})$ with $0$ and $c_i$ with $0$ for $i \geq m$.
\State \hspace{2ex} Initialize $\gamma_i$ with $0$ for $i \in \{0, 1, \dots, m-1\}$.
\State \hspace{2ex} \textbf{for} $i \in \{m-1, m-2, \dots, 0\}$ \textbf{do} \textit{(in order)}
\State \hspace{4ex} \textbf{for} $k_i \in \{1, 2, \dots, x\}$ \textbf{do}
\State \hspace{6ex} \textbf{if} $c_{i+k_i} = \alpha^{q-2}$ \textbf{then}
\State \hspace{8ex} Set $\gamma_i = x-k_i+1$.
\State \hspace{8ex} Set $\textup{index} = i - \gamma_i$.
\State \hspace{8ex} \textbf{break}. \textit{(exit current loop)}
\State \hspace{6ex} \textbf{end if}
\State \hspace{4ex} \textbf{end for}

\State \hspace{4ex} \textbf{if} $c_i \neq 0$ \textbf{then} \textit{(same as $a_i \neq 0$)}
\State \hspace{6ex} Set $a_i = \mathcal{L}(c_i)$.
\State \hspace{6ex} $g(\bold{c}) \leftarrow g(\bold{c}) + a_i (q-1)^{\gamma_i} N_q(\textup{index}, x)$.
\State \hspace{4ex} \textbf{end if}
\State \hspace{2ex} \textbf{end for}
\State \hspace{2ex} Compute $\bold{b}=\textup{binary}(g(\bold{c})-1)$, which has length $s^{\textup{c}}$.
\State \hspace{2ex} Ignore the next $x$ bridging symbols.
\State \textbf{end for}
\State \textbf{Output:} Outgoing stream of binary messages.
\end{algorithmic}
\label{alg_dec}
\end{algorithm}

In order to reduce complexity, all terms containing multiplications in Algorithms~\ref{alg_enc} and \ref{alg_dec}, e.g., $a_i (q-1)^{\gamma_i} N_q(\textup{index}, x)$, are not computed at runtime. This increases the storage overhead, which will be discussed shortly. However, the gain is that the complexity of both algorithms is still mainly governed by the adder size that will perform the comparisons/subtractions and additions. The adder size is itself the message length $s^\textup{c}$. For example, to achieve a rate of $1.8519$ information bits per coded symbol using a CQA-LOCO code with $q=4$ and $x=1$, adders of size $1.8519 \times (26+1) = 50$ bits are needed (see Table~\ref{table1}). Another example is, to achieve a rate of $1.8000$ information bits per coded symbol using a CQA-LOCO code with $q=4$ and $x=2$, adders of size $1.800 \times (38+2) = 72$ bits are needed (see Table~\ref{table2}).

As illustrated in the previous paragraph, the storage overhead increases as $q$ increases. In particular, and from Steps~18--31 in Algorithm~\ref{alg_enc} and Steps~14--17 in Algorithm~\ref{alg_dec}, the storage grows with $O((q-1)x\log_2 q)$, $q > 2$, for fixed $m$. The term $\log_2 q$ is there because the storage needed for cardinalities only,~which are computed offline, grows with $O(\log_2 q)$ for fixed $m$. Moreover, from Steps~18--31 in Algorithm~\ref{alg_enc} (resp., Steps~14--17 in Algorithm~\ref{alg_dec}), the encoding complexity (resp., decoding complexity) grows with $O((q-1)\log_2 q)$ (resp., $O(\log_2 q)$) for fixed $m$. The term $\log_2 q$ is there because the adder size grows with $O(\log_2 q)$ for fixed $m$ as implied in the examples of the previous paragraph.

However, these orders of growth result in an unfair comparison across different values of $q$ because they are based on a fixed number of symbols rather than the same amount of coded data. For example, if $m$ is fixed at $25$, these are $25$ bits for $q=2$, but equivalent to $50$ bits for $q=4$, to $75$ bits for $q=8$, and so on. Thus, these orders of growth should be divided by $\log_2 q$ for a fair comparison, which results in $O((q-1)x)$ for storage, $O(q-1)$ for encoding complexity, and $O(1)$ for decoding complexity.\footnote{We can also choose to fix the message length in bits instead of fixing the amount of coded data. Note that while CQA-LOCO codes with higher values of $q$ have higher rates, the effect of this on the \textit{orders} of growth is minor.} Thus, the storage and complexity of QA-LOCO encoding and decoding with $q > 2$ are still manageable, and are less than other enumerative techniques. One useful comparison to make is against the complexity growth of non-binary low-density parity-check (NB-LDPC) decoding, which has $O(q \log_2 q)$ that goes down to $O(q)$. The QA-LOCO order of storage and encoding-complexity growth is quite nearly $O(q)$, and the QA-LOCO order of decoding-complexity growth is even much better.

A Flash device with $q$ levels per cell has $\log_2 q$ pages. In general, the Flash industry prefers to process different pages independently in order to reduce latency. One idea to achieve this goal is to apply the QA-LOCO code only on the parity part of the component LDPC code as we did in \cite{ahh_loco} for MR systems. In particular, the idea is to group the parity bits of $\log_2 q$ LDPC codewords that have their information bits to be written over the available $\log_2 q$ different Flash pages; one codeword per page, convert these parity bits into symbols over GF($q$), and encode them via a QA-LOCO code before writing them; one symbol per cell. While reading, the parity bits are decoded via the QA-LOCO decoder first, and then the LDPC decoder operates independently on the $\log_2 q$ pages to retrieve the $\log_2 q$ codewords. High performance LDPC codes for Flash can be designed according to \cite{ahh_jsac}, \cite{ahh_nboo}, and \cite{homa_boo}.

The fact that the encoding and decoding of QA-LOCO codes are performed through simple adders enables reconfigurability. All that is needed to reconfigure a QA-LOCO code, i.e., change the code parameters such that more (or even different) constraints are supported, is to change the cardinalities that are inputs to the adders at both encoding and decoding sides such that the encoding-decoding rule in (\ref{eqn_rule}) supports the new constraints. As the Flash device ages, charges propagate during programming with higher rates and to further non-adjacent cells. Thus, while QA-LOCO codes with $x=1$ are sufficient when the device is fresh, reconfiguring to QA-LOCO codes with $x > 1$, i.e., forbidding more patterns, is needed such that the device keeps functioning reliably late in its lifetime.

Aided by machine learning, errors before the LDPC decoder can be collected to identify the set of error-prone patterns that should be forbidden at different stages of the Flash device lifetime. Once this set is found to be bigger that the currently supported set by the QA-LOCO code, we propose to respond via reconfiguring the QA-LOCO code to support the new set as illustrated in the previous paragraph. Therefore, machine learning and reconfigurable constrained codes can help increase the lifetime of modern Flash devices significantly, and therefore support the evolution of QLC and PLC Flash memories.

\vspace{+0.3em}
\section{Conclusion}\label{sec_conc}

We introduced capacity-achieving $q$-ary asymmetric LOCO codes (QA-LOCO codes) for Flash devices with any number, $q$, of levels per cell. We partitioned the codewords of a~QA-LOCO code into groups, which we used to recursively compute the cardinality. We devised an encoding-decoding rule for QA-LOCO codes to map from index to codeword~and vice versa, which is the key result behind the simple encoding and decoding of these codes. We introduced the achievable rates of QA-LOCO codes, and showed that they need $5\%$ or less redundancy to protect the device. For QLC and PLC devices, we demonstrated that ICI mitigation almost comes for free with respect to redundancy. We presented the encoding and decoding algorithms, and provided an analysis for the storage and complexity growth with $q$. We suggest that machine learning and reconfigurable QA-LOCO codes can significantly increase the lifetime of modern Flash devices.


\end{document}